\documentclass[amsmath,amssymb,twocolumn,notitlepage,nofootinbib,superscriptaddress,floatfix,eprint]{revtex4-2}

\usepackage[utf8]{inputenc}
\usepackage{graphicx}
\usepackage{braket}
\usepackage{multirow}
\usepackage{booktabs}
\usepackage{hyperref}
\usepackage{xcolor}
\usepackage{float}
\usepackage{subcaption}
\usepackage{caption}
\usepackage{amsthm}
\usepackage{dsfont}

\theoremstyle{plain}

\theoremstyle{plain}

\theoremstyle{plain}

\theoremstyle{remark}
\newtheorem*{rem*}{\protect\remarkname}
\theoremstyle{plain}

\theoremstyle{plain}

\theoremstyle{definition}

\theoremstyle{plain}
\newtheorem*{thm*}{\protect\theoremname}
\theoremstyle{plain}
\newtheorem*{lem*}{\protect\lemmaname}

\providecommand{\propositionname}{Proposition}
\providecommand{\theoremname}{Theorem}
\providecommand{\lemmaname}{Lemma}
\providecommand{\remarkname}{Remark}
\providecommand{\conjecturename}{Conjecture}
\providecommand{\definitionname}{Definition}
\providecommand{\corollaryname}{Corollary}
\allowdisplaybreaks

\begin{document}
\title{Experimental data re-uploading with provable enhanced learning capabilities}

\author{Martin F. X. Mauser}
\email{martin.mauser@univie.ac.at}
\affiliation{University of Vienna, Faculty of Physics, Vienna Center for Quantum
Science and Technology (VCQ), Boltzmanngasse 5, Vienna 1090, Austria}
\affiliation{University of Vienna, Vienna Doctoral School in Physics,  Boltzmanngasse 5, Vienna 1090, Austria}

\author{Solène Four}
\affiliation{University of Vienna, Faculty of Physics, Vienna Center for Quantum
Science and Technology (VCQ), Boltzmanngasse 5, Vienna 1090, Austria}
\affiliation{Master Quantum Engineering, Ecole Normale Supérieure, Université PSL, 75005 Paris, France }

\author{Lena Marie Predl}
\affiliation{University of Vienna, Faculty of Physics, Vienna Center for Quantum
Science and Technology (VCQ), Boltzmanngasse 5, Vienna 1090, Austria}

\author{Riccardo Albiero}
\affiliation{Istituto di Fotonica e Nanotecnologie, Consiglio Nazionale delle Ricerche (IFN-CNR), piazza L. Da Vinci 32, 20133 Milano, Italy}

\author{Francesco Ceccarelli}
\affiliation{Istituto di Fotonica e Nanotecnologie, Consiglio Nazionale delle Ricerche (IFN-CNR), piazza L. Da Vinci 32, 20133 Milano, Italy}

\author{Roberto Osellame}
\affiliation{Istituto di Fotonica e Nanotecnologie, Consiglio Nazionale delle Ricerche (IFN-CNR), piazza L. Da Vinci 32, 20133 Milano, Italy}

\author{Philipp Petersen}
\affiliation{University of Vienna, Faculty of Mathematics and Research Network Data Science @ Uni Vienna, Kolingasse 14-16, Vienna 1090, Austria}

\author{Borivoje Daki\'{c}}
\affiliation{University of Vienna, Faculty of Physics, Vienna Center for Quantum
Science and Technology (VCQ), Boltzmanngasse 5, Vienna 1090, Austria}
\affiliation{Institute for Quantum Optics and Quantum Information Sciences (IQOQI), Austrian Academy of Sciences, Boltzmanngasse 3, Vienna 1090, Austria}
\affiliation{QUBO Technology GmbH, Vienna 1090, Austria}

\author{Iris Agresti}
\email{iris.agresti@univie.ac.at}
\affiliation{University of Vienna, Faculty of Physics, Vienna Center for Quantum
Science and Technology (VCQ), Boltzmanngasse 5, Vienna 1090, Austria}

\author{Philip Walther}
\email{philip.walther@univie.ac.at}
\affiliation{University of Vienna, Faculty of Physics, Vienna Center for Quantum
Science and Technology (VCQ), Boltzmanngasse 5, Vienna 1090, Austria}
\affiliation{Institute for Quantum Optics and Quantum Information Sciences (IQOQI), Austrian Academy of Sciences, Boltzmanngasse 3, Vienna 1090, Austria}
\affiliation{QUBO Technology GmbH, Vienna 1090, Austria}

\begin{abstract}
The last decades have seen the development of quantum machine learning, stemming from the intersection of quantum computing and machine learning. This field is particularly promising for the design of alternative quantum (or quantum inspired) computation paradigms that could require fewer resources with respect to standard ones, e.g. in terms of energy consumption. In this context, we present the implementation of a data re-uploading scheme on a photonic integrated processor, achieving high accuracies in several image classification tasks. We thoroughly investigate the capabilities of this apparently simple model, which relies on the evolution of one-qubit states, by providing an analytical proof that our implementation is a universal classifier and an effective learner, capable of generalizing to new, unknown data. Hence, our results not only demonstrate data re-uploading in a potentially resource-efficient optical implementation but also provide new theoretical insight into this algorithm, its trainability, and generalizability properties. This lays the groundwork for developing more resource-efficient machine learning algorithms, leveraging our scheme as a subroutine.
\end{abstract}

\maketitle

\section{Introduction}
The last decades have witnessed the emergence of quantum computing as a new paradigm to solve problems, with the promise of outperforming standard methods. In particular, the discovery of Shor's and Grover's algorithms \cite{shor1994algorithms, grover1996fast} has ultimately proven that quantum computers can efficiently tackle computational problems that are hard for any known classical algorithm. However, tasks with a wide range of applications, like prime numbers factorization are still out of reach for the current state-of-the-art of quantum hardware, and a quantum advantage has been only demonstrated for problems with no practical use, like Boson Sampling and Random Circuit Sampling \cite{aaronson2011computational, lund2017quantum,arute2019quantum,zhong2020quantum,madsen2022quantum}. On the other hand, a research field that has attracted a flurry of interest stems from the combination of quantum computing with machine learning \cite{biamonte2017quantum, dunjko2018machine}. In this context, the aim has been to create a fruitful intersection, where quantum computers can provide a boost with applicative implications, by enhancing the performance of standard machine learning algorithms. Mutually, the latter can offer methods to gain deeper insights on quantum systems, either by post-processing large amount of data or by directly dealing with quantum data without the need of classical description.


\begin{figure*}[tb]
    \centering   \includegraphics[width=0.8\textwidth]{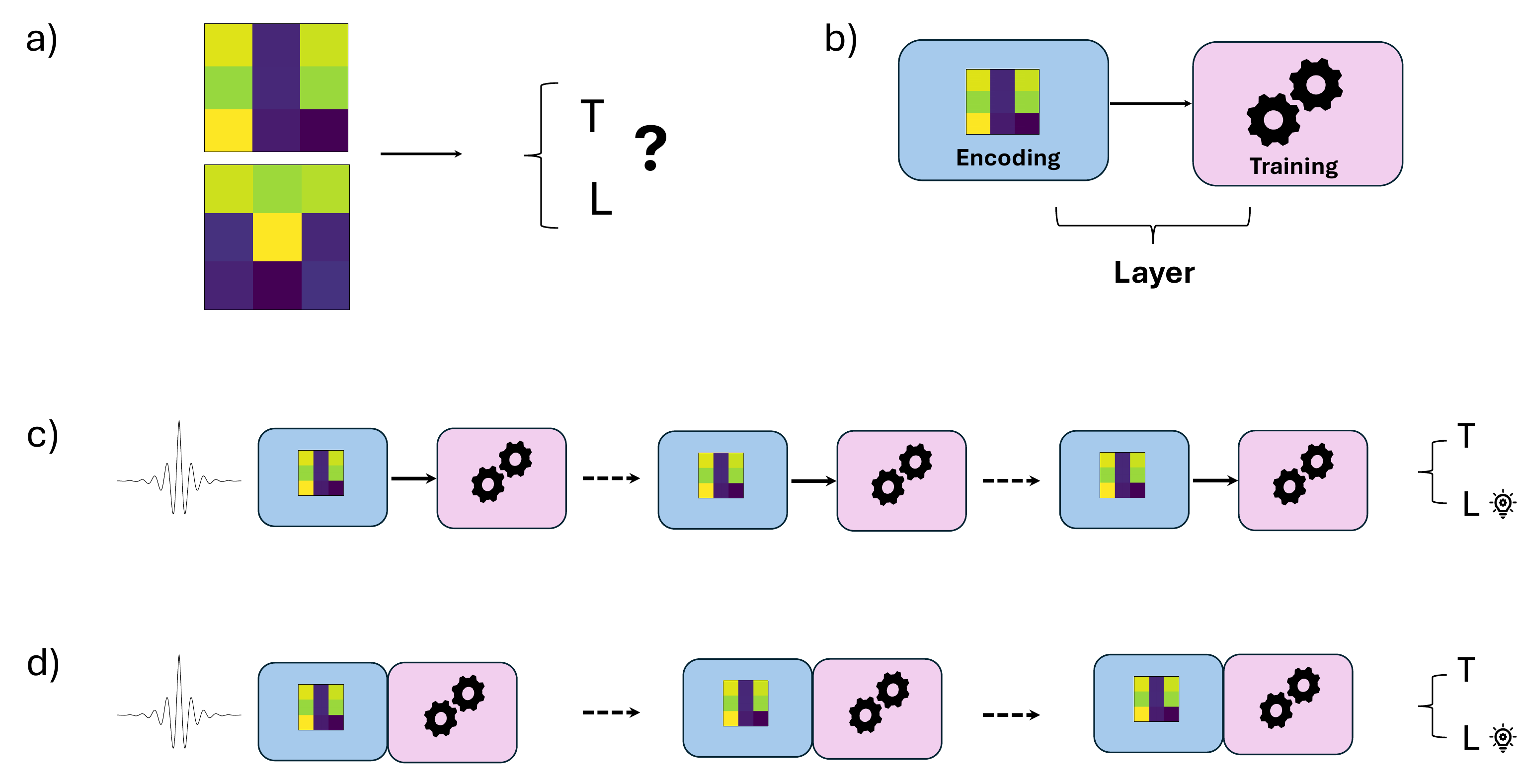}
    \caption{\textbf{Image classification through data re-uploading.} \textbf{a)} The classification tasks we tackle in this work amount to assign a given image to one, out of two, classes. Here we report the example of tetromino images, picturing either a ``L" or a ``T" and our model needs to understand which letter is represented. \textbf{b)} The data re-uploading algorithm is composed by the sequence of many layers. One layer is composed by the encoding of the input data (in this case the image we want to classify) in a qubit state, followed by a random rotation, whose parameters are optimized through a training phase. \textbf{c)} The initial qubit state, pictorially represented by a wavepacket, which is fixed, is injected into a sequence of many layers (alternating encoding and rotations). The optimal rotations are found after a training phase, ensuring that the output state holds the information of whether the input belongs to the first or second class. In the original proposal, the encoding and the rotations are implemented in two separate gates and we refer to this scheme as \textit{original}. \textbf{d)} A variant of the algorithm merges the encoding and the rotation in the same gate. On one side, this reduces the length of the circuit, but, on the other, it hinders the generalizability of the method. We refer to this scheme as \textit{compressed}. }
    \label{fig:conc_scheme}
\end{figure*}

Considering the challenges involved in the implementation and control of a high number of qubits and the possible experimental errors, it is crucial to identify the lowest amount of quantum resources, which allows us to tackle relevant tasks. The answer to this question was recently offered by the development of a quantum machine learning algorithm called \textit{data re-uploading} \cite{Perez-Salinas_2020}, whose concept is depicted in Fig. \ref{fig:conc_scheme}. 
Despite its apparent simplicity, this model achieves highly complex mappings of the input data, constituting an alternative to kernel methods \cite{schuld2021quantum, liu2021rigorous, huang2021power, Jerbi_2023}, 
and it has been proven to be a \textit{universal approximator} \cite{Perez-Salinas_2020, Goto_2021}. Furthermore, the re-uploading of the input data circumvents the \textit{no cloning theorem} \cite{wootters1982single}, which prevents quantum models from easily copying and retrieving the input multiple times, as it would happen in the classical case. 
This model has been widely investigated from the theoretical point of view \cite{Schuld_2021, Goto_2021, Yu_2022, jerbi2023quantum} and it was also implemented on several quantum platforms \cite{Ono_2023,  Perez-Salinas_2021, Fan_2022, Wach_2023, Easom_2021}.
However, most of the reported experiments slightly diverge from the theoretical proposal, since they express each encoding of the input data followed by a qubit rotation as a single gate (we will refer to this model as \textit{compressed}). The two schemes, i.e. the original and the compressed one, are shown in Fig.~\ref{fig:conc_scheme}c and Fig.~\ref{fig:conc_scheme}d respectively. Although this might seem a minor or even non-existing difference, it has a huge impact on the complexity of the learning model and it drastically affects its generalisability and trainability \cite{Caro_2021, Chen_2022, Gil-Fuster_2024}. 
To properly understand this concept, let us consider that the quality of a learner is defined by two aspects: its ability to remember labelled inputs and how well it can generalise its operation to new unknown data.
Hence, choosing the best learning model will be a balancing act between an overly expressive and slightly underexpressive one, i.e. the complexity of the learner should match the complexity of the task. 

\begin{figure}[htbp!]
    \centering   \includegraphics[scale=0.55]{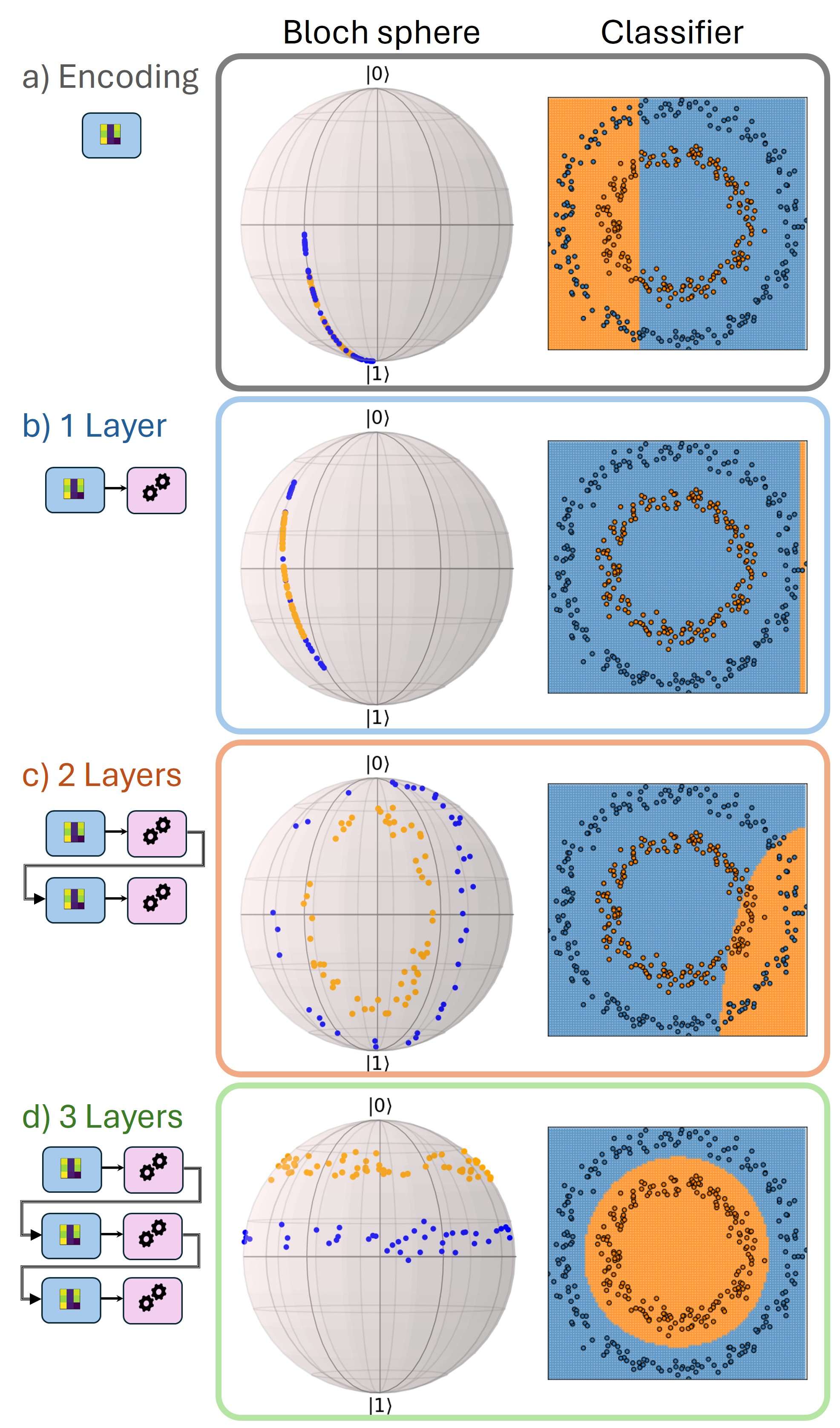}
    \caption{\textbf{Photonic data re-uploading.} We show the working principle of our photonic data re-uploading algorithm, illustrating the example of the circles dataset, where we want our model to distinguish between inner points and external ones.
    We start with the encoding of the data points (\textbf{a}) and we show the effect of this first gate on the initial state on the Bloch sphere. The blue (yellow) color represents the correct label of the point. On the right, we show the prediction of the algorithm if we would stop the algorithm here, gaining a poor performance. Again, the blue (yellow) color of the dot represents the true label, while the background shows the label predicted by the model. In other words, all the points in the yellow (blue) region are classified as yellow (blue). Then, we add the first rotation (\textbf{b}) and we show the action of a second (\textbf{c}) and third (\textbf{d}) layer. It is visible from the shape of the coloured regions in the background how, increasing the number of layers, the model gets more complex and nonlinear. In this case, already after 3 layers, the algorithm successfully classifies the circles dataset.} 
    \label{fig:qubit_scheme}
\end{figure}

In this work, we demonstrate a realization of a data re-uploading scheme on a tunable integrated photonic processor \cite{pentangeloHighfidelityPolarizationinsensitiveUniversal2024} fabricated by femtosecond laser waveguide writing \cite{corrielliFemtosecondLaserMicromachining2021}, which, due to its architecture, ensures that the algorithm is designed according to the initial theoretical proposal, i.e. separating the input encoding from the optimizable part. We show its working principle in Fig.~\ref{fig:qubit_scheme}. From a theoretical point of view, we demonstrate its universality (complementing the recent work \cite{perez2024universal}), and prove analytically that our model succeeds in tasks in which previous implementations fall short, while also showcasing better convergence to the optimal parameters, during the training phase.
Furthermore, from the experimental point of view, we show its performance on a vast set of binary image classification tasks of increasing complexity. To reach this result, we exploit two kinds of input states: first, single photon states, to prove that our model can be used as a building block for more complex quantum computing architectures, and, second, coherent light, which could indicate a possible path towards scalable optical computing models. These results offer deep insights on the generalizability of quantum machine learning models and show how quantum architecture can offer inspiration to formulate optimization problems in a more convenient way. This work opens also the way to optical computing models, which promise to bring more energy efficient algorithm executions \cite{hamerly2019large}.


\begin{figure*}[tb]
    \centering   \includegraphics[width=\textwidth]{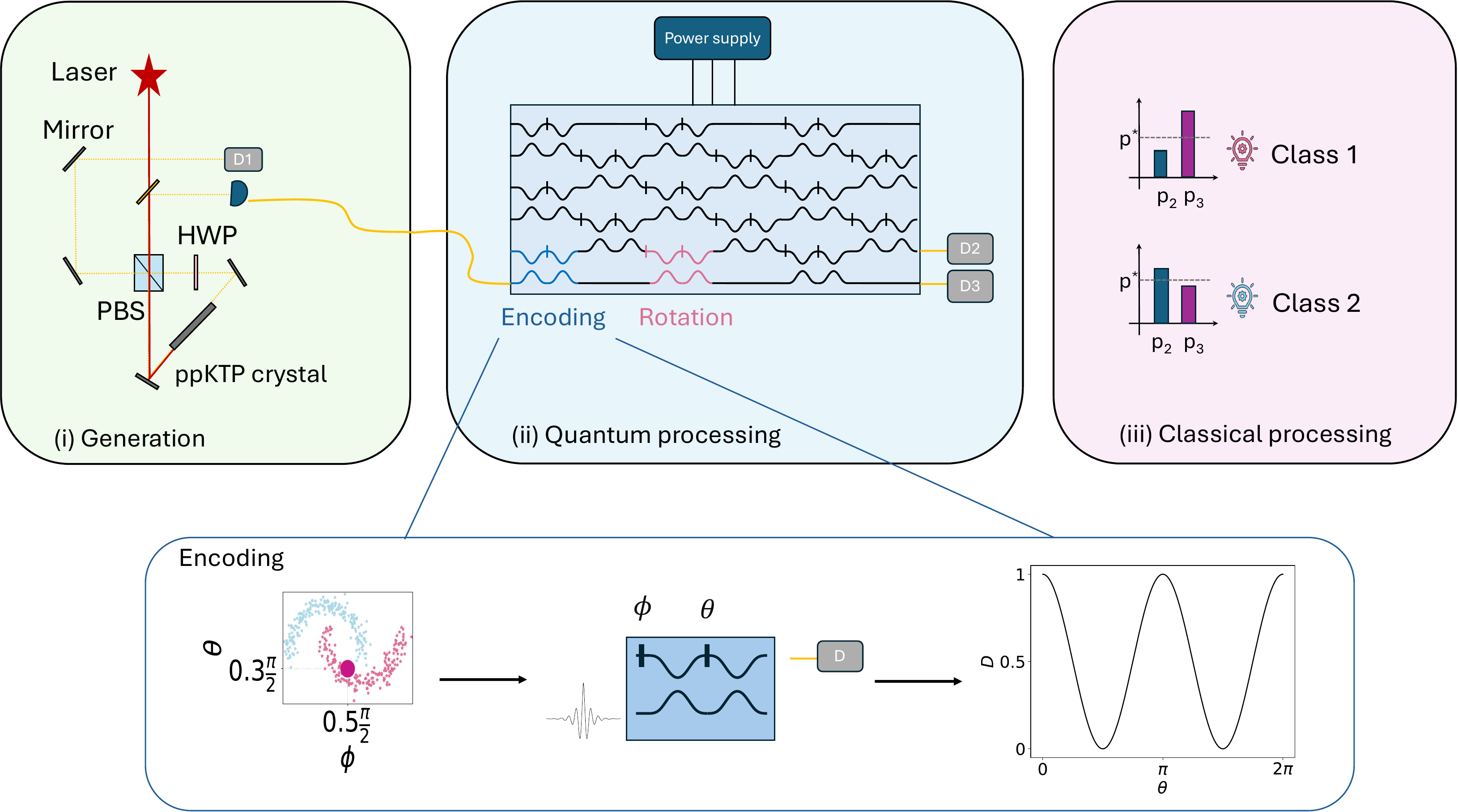}
    \caption{\textbf{Experimental apparatus.} Our algorithm is constituted of three main parts: (i) the generation of the single photons encoding the fixed qubit state to be evolved, (ii) the quantum processing of the state, and (iii) the classical processing, which performs the final classification. 
    (i) The input of the model featuring single photon states is generated through a Spontaneous Parametric Down-Conversion source, constituted by a periodically poled Potassium titanyl-phosphate crystal (KTP), pumped by a $775~\mathrm{nm}$ laser. The crystal is pumped in one direction and a separable state is generated, where one photon is used as trigger and the second one is sent in the circuit. The latter encodes one qubit in the path degree of freedom, through the dual-rail scheme.
    (ii) Then, the quantum processing, namely the controlled evolution of such a qubit state, is performed through an integrated photonic circuit, whose operation is tunable, using phase shifters. In particular, one layer of the data re-uploading scheme is constituted by the sequence of two tunable Mach-Zehnder interferometers (MZIs), where one is exploited to encode the input data and the second to rotate it. 
    In the inset, we show the encoding procedure, through the example of the moon dataset, where we want to discriminate between the purple and blue points. Here, since each data point has 2 coordinates ($x_1$, $x_2$), the encoding will consist in evolving our input state through the action of a MZI, whose external phase will amount to $x_1\frac{\pi}{2}$ and internal one will be $x_2\frac{\pi}{2}$. As shown in the graph, this gives a nonlinear transformation of the input. At the end, the output mode from where the photon leaves the circuit is recorded, through single photon detectors. Let us note that we post-select our statistics, considering only the events where one of the two detectors at the output of our circuit and the trigger click simultaneously. (iii) The third step is fully classical and it consists in taking the output statistics and performing a Linear Discriminant Analysis (LDA) algorithm, that sets a threshold on one of the probabilities. If this threshold is overcome, the point is assigned to class 1, otherwise to class 2. The training phase can be either simulated, until the optimal parameters are found and implemented, or performed on-chip, through a parameter shift algorithm \cite{Schuld_2019}.}
    \label{fig:experimental_apparatus}
\end{figure*}

\section{Photonic data re-uploading}
The model that we design and implement in this work follows the original proposal of data re-uploading \cite{Perez-Salinas_2020}.
Its structure amounts to consecutive rotations $U(\theta_1, \theta_2, \theta_3)$ applied to a single qubit input, which we denote as $|\Psi_{in}\rangle$. Conceptually, in this sequence, we can distinguish two kinds of rotations: (i) \textit{encoding} and (ii) \textit{processing}. The first is required to upload the input data to be classified into the circuit; the second, instead, has the scope of implementing a non-trivial feature map of the input into an output state $|\Psi_{out}\rangle$, so that a linear classification is sufficient in the end to perform the desired task.

\begin{figure*}[tb]
    \includegraphics[width=\textwidth]{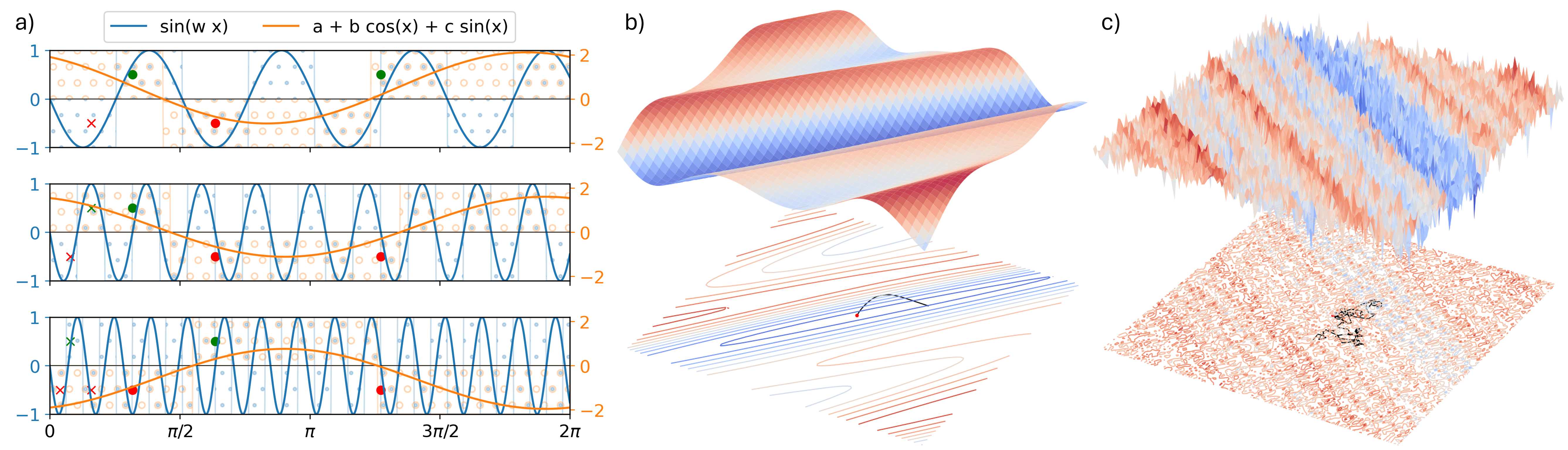}
    \caption{\textbf{Generalizability and Trainability of data re-uploading models.} \textbf{a)} Visual representation of how the compressed data re-uploading scheme (indicated with the blue curve) can correctly label arbitrarily many training data points, while the original one (studied in this work) can only do the same for up to 3 points. This implies that the compressed scheme has an infinite VC dimension, while ours is finite, i.e. $VCdim = 3$ (for one layer). This ensures that our model can generalize to test data points. The marked points indicate the training dataset, while the color and vertical position indicate the true label. The shaded areas indicate the classification result for the two learning models. The shape of the marked points indicate if both models where able to classify all points at that horizontal position correctly (dot) or only the blue sinusoidal classifier with tunable frequency (cross). \textbf{b)} and \textbf{c)} picture the loss landscape in the two cases: original (\textbf{b}) and compressed (\textbf{c}). Two layers are considered for both images and the loss function has been projected to 2 dimensions via a principal component analysis for visualization purposes. This demonstrates how the trainability of the model is influenced by its VC dimension. In addition, one possible optimization path using the Adam optimizer is shown on the lower heatmap of the two loss landscapes.}
    \label{fig:vis_VCdim}
\end{figure*}

\begin{figure*}[tb]
    \centering   \includegraphics[width=0.8\textwidth]{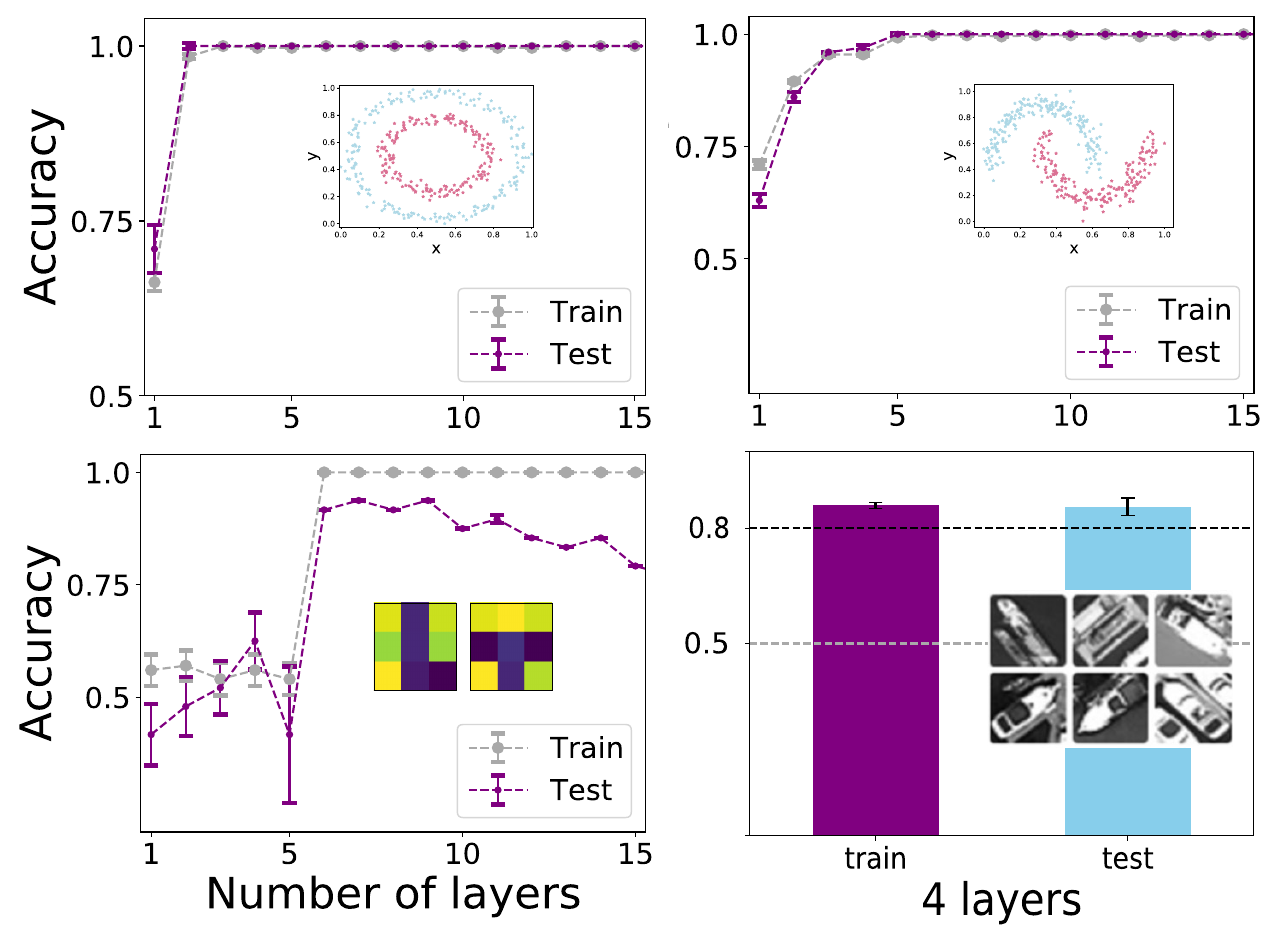}
    \caption{\textbf{Experimental classification accuracies.} We report the classification accuracies obtained through our photonic data re-uploading model for the classification of four classes of images: (i) circles, (ii) moons, (iii) tetromino and (iv) images from the overhead MNIST dataset, representing ships and cars. The grey dots represent the training accuracy, while the purple ones the testing. The first two sets are composed of bi-dimensional data (which are the $(x,y)$ coordinates on the plane), so the base layer block is only composed of 2 MZIs. In this case, we used 400 samples as training and 100 as test. The tetromino images, instead, are composed of 9 pixels (containing a number between 0 and 1), implying that we need a base layer block of 10 MZIs. In this case, we used 100 noisy data samples for training and 48 noiseless for test. Regarding the Overhead MNIST images, instead, they were composed of $28 \times 28$ pixels, reduced to $20$ parameters through principal component analysis \cite{jolliffePrincipalComponentAnalysis1986,Pearson01111901}. Hence the base layer block amounts to  20 MZIs. In this case, we used 1776 data samples for training and 222 for test. For the cases (i) and (ii), it can be seen that the accuracy in the prediction gets better when increasing the number of layers, reaching the maximum after 3 layers. For the tetromino dataset, we see a drop in the test accuracy, due to overfitting when the number of layers (or free parameters) is too high. In the case of the Overhead MNIST images, we report the accuracy obtained for 4 layers. The error bars are obtained through a Monte Carlo simulation, considering 1000 repetitions for each training and testing, where we consider that the underlying statistics for our photon detection is Poissonian, and represent one standard deviation.}
    \label{fig:results}
\end{figure*}
Hence, from a mathematical point of view, the overall operation of the model is the following:

\begin{equation} 
\label{eq:Data_re-uploading}
    \ket{\Psi_{out}} = \mathcal{U}(\vec{x}, \vec{\xi})\ket{\Psi_{in}}=\prod_{l=1} ^L [U_l(\vec{\theta})U(\vec{x})] \ket{\Psi_{in}}
\end{equation}

where $\vec{x}$ represents the initial features of the classical input data, which will define all of the encoding unitaries, $\vec{\xi}=(\vec{\theta_1}, ..., \vec{\theta_L})$ is a vector containing the optimizable parameters, which will amount to the processing rotations. Then, a pair $U_l(\vec{\theta}) U(\vec{x})$ composes a \textit{layer} of the circuit (see Fig.~\ref{fig:conc_scheme}b). Hence, for $L$ layers, we will have a total circuit depth, or number of gates, of $2L$.
In the end, the output state is projected on the computational basis ($|0\rangle$, $|1\rangle$), constituted by the eigenvectors of the Pauli $Z$ operator. The result will be a bi-dimensional vector $(p, 1-p)$, where $p=Tr(|0\rangle \langle0|\Psi_{out}\rangle \langle \Psi_{out}|)$. This data is then classically post-processed through Linear Discriminant Analysis (LDA) \cite{hastie2005elements}, to find the linear separation between the two considered classes. 

This implies that the classical algorithm will find a threshold $\tau$, such that, if $p \geq \tau$, with
\begin{equation} 
\label{eq:Data_re-uploading}
p=Tr(|0\rangle \langle0|\mathcal{U}(\vec{x}, \vec{\xi})|\Psi_{in}\rangle \langle \Psi_{in}|\mathcal{U^\dagger}(\vec{x}, \vec{\xi}))
\end{equation}

the data point $\vec{x}$ will be assigned to class $1$, otherwise to class $2$.

Analogously to standard artificial neural networks, the higher the number of layers the circuit will display, the better its representation capabilities will be, implying an overall more powerful classifier \cite{Goto_2021}, as it is also visible from Fig.~\ref{fig:qubit_scheme}. While this might constitute an issue for platforms like superconducting qubits, due to their limited coherence time, it is particularly suitable for photonic implementations.

Let us consider that, although in this model the input data points are introduced linearly in the rotations, the overall feature map will be anyway non-linear, due to the structure of these gates, ensuring that it can potentially implement a universal classifier \cite{Schuld_2021, Goto_2021, Yu_2022}.
To show such a non-linearity, let us consider that in our model each layer of the classifier is implemented by a pair of Mach-Zehnder interferometers (MZI), operated in the path degree of freedom see Fig.~\ref{fig:experimental_apparatus}. Such a gate can implement any rotation of the Bloch sphere and it features two external phases ($\phi_1$, $\phi_2$) and an internal one ($\theta$) and is represented through the following Jones matrix:

\begin{equation}
\label{mzi_form}
ie^{i\frac{\theta}{2}}
\begin{pmatrix}
e^{i(\phi_1+\phi_2)} \mathrm{sin}(\frac{\theta}{2}) & e^{i(\phi_1)} \mathrm{cos}(\frac{\theta}{2}) \\
e^{i(\phi_2)}\mathrm{cos}(\frac{\theta}{2}) & -\mathrm{sin}(\frac{\theta}{2})
\end{pmatrix}
\begin{pmatrix}
0 \\
1
\end{pmatrix}
\end{equation}

where we inject the input state $|1\rangle$, which, in the dual-rail encoding, corresponds to the presence of a photon (or of a coherent light input) on one of the spatial modes of the MZI, see Fig.~\ref{fig:experimental_apparatus}. From Eq.~\eqref{mzi_form}, it is visible that a projective measurement on the $Z$ basis will produce probabilities that depend non-linearly on the phases. Let us note that, for deriving Eq.~\eqref{mzi_form}, we considered the dielectric beam-splitter.

However, since our input is not a superposition of $|0\rangle$ and $|1\rangle$, we need one less degree of freedom and we can consider $\phi_2=0$. So, for the sake of notation, for the rest of the paper, we will omit the subscripts and indicate $\phi_1$ simply as $\phi$. For the same reason, let us note that we are not sensitive to the $\phi_1$ of the first Mach-Zehnder of our circuit, as it can be seen from Eq.~\eqref{mzi_form}.

Then, to encode our data, assuming that it is constituted by bi-dimensional points $x=\{x_1, x_2\}$, we use the following map: $\phi = x_1 \frac{\pi}{2}$ and $\theta = x_2 \frac{\pi}{2}$ (see Fig.~\ref{fig:experimental_apparatus}). 
In the case of higher dimensional input data, e.g. $x=\{x_1, ..., x_n\}$, if $n$ is an even number, we proceed by encoding the coordinates in consecutive unitaries, in the following way:
\begin{equation}
\label{encoding_scheme}
    \Pi_{j=0}^{ \lceil\frac{n}{2}\rceil-1} U(\theta_{2j}, \theta_{2j+1}) U(x_{2j}, x_{2j+1})
\end{equation}

Instead, if $n$ is odd, an extra layer needs to be added, implementing $U(\theta_{n}, \theta_{n+1}) U(x_{n}, 0)$.
Hence, the lowest number of gates required to encode classical data of dimension $n$ amounts to $\frac{n}{2}$ and the size of the array of trainable parameters is at least as big as one of the input. We will refer to the building block of our circuit, reported in Eq.~\eqref{encoding_scheme} as \textit{base layer block} \cite{Periyasamy_2022}.

\section{Trainability, generalizability and universality of the model}

Similar to other supervised machine learning models, the performance of data re-uploading amounts to the accuracy with which the algorithm can reproduce the labelling of training data and of test data, which is unknown. Although both are significant, the latter is more relevant, since the main purpose of machine learning models is to deal with unknown inputs, namely they should have good generalization properties.
Several tools have been introduced to quantify generalization \cite{Caro_2021, Chen_2022, Gil-Fuster_2024}, but one of the most important 
is the Vapnik-Chervonenkis (VC) dimension \cite{vapnik2015uniform, Petersen_2024, Shalev-Shwartz_2014}. This parameter is tightly related to the complexity and expressivity of the model itself.

More formally, the VC dimension of a class of functions \( \mathcal{H} \) (in our case those that our classifier can implement) is defined as the size of the largest set of points for which, for every possible assignment of labels (0 or 1), there exists a function in \( \mathcal{H} \) that classifies them correctly.
If we now consider the theoretical framework of Probably Approximately Correct (PAC) learning \cite{valiantTheoryLearnable1984} the minimum amount of data required to achieve a given error rate is proportional to the VC dimension. This implies that, for a model featuring an infinite VC dimension, no finite amount of data suffices to correctly classify unknown data.  
Moreover, another side effect of high complexity is that the training procedure could be extremely inefficient or even impossible due to a very irregular optimization landscape for the loss function \cite{liVisualizingLossLandscape2018}. Hence, for any task, a trade-off must be found between the expressivity of the model and its generalizability.

To mathematically show the implications of the different VC dimensions in the two schemes, let us consider the functions that a single layer can implement in each of the two models. In the compressed case, the functions that can be implemented have the following form (see Supplementary Material note I for derivation):

\begin{equation}
\label{compressed_model}
    \mathcal{H} = \{x\mapsto \sin{(\omega x)}| \omega \in \mathbb{R}\}
\end{equation}

and it can be shown that $\mathcal{H}$ has an infinite VC dimension. This is pictorially represented in Fig.~\ref{fig:vis_VCdim}, where we consider a toy model, where the datapoints are given as $X_i = 2^i$, and $i$ represents an index ranging from 0 to $N$. The corresponding labels are drawn randomly from $y_i \in [0, 1]$. It is visible that, with the model of Eq.~\eqref{compressed_model}, it is possible to reproduce any labelling choosing the following frequency $\omega$:
\begin{equation}
\label{eg:frequency_ToyModel}
    \omega = 2\pi \sum_{i=0} ^N \frac{y_i}{2^i}.
\end{equation} 

This proves that the VC dimension of this learning model is in fact infinite, as it was also demonstrated in{mohriFoundationsMachineLearning}.

On the contrary, the original scheme, when featuring a single layer, can only reproduce the following set of functions:
\begin{equation}
\label{our_model_1}
\mathcal{H} = \{x \mapsto \text{sign}[a+b\cos{(x)} + c\sin{(x)}]| a,b,c \in \mathbb{R}\}.
\end{equation}
which can be described by the following periodic interval classifier:

\begin{equation}
\label{our_model_2}
\mathcal{H} = \{x \mapsto \mathds{1}_{\mathcal{S}=[a + 2\pi n, b + 2\pi n]} | a,b \in \mathbb{R}; \forall n \in\mathbb{Z}\}.
\end{equation}

This implies that all the points in the interval \mbox{$[a + 2\pi n, b + 2\pi n]$} are classified as belonging to class 1 and those outside to class 2. Both Eq.~\eqref{our_model_1} and Eq.~\eqref{our_model_2} display $VCdim(\mathcal{H}) = 3$ (see Supplementary Material note I). 
Hence, as it is shown in Fig.~\ref{fig:vis_VCdim}, the model can correctly reproduce arbitrary labellings of only 3 points.
Then, when considering two layers, the interval $\mathcal{S}$ becomes $[a + 2\pi n, b + 2\pi n] \cup [c + \pi n, d + \pi n]$ with $a,b,c,d \in \mathbb{R}$ and $\forall n \in \mathbb{Z}$, with a corresponding VC dimension $VCdim(\mathcal{H}) = 5$. In general, it holds that $\ell$ layers lead to a VC dimension of $2\ell+1$. Concretely, we can choose $2\ell+1$ distinct points $x_1, \dots, x_{2\ell+1}$ in an interval $[a,b]$ and show that they can be completely separated according to their label (\textit{shattered}) by $\ell$ periodic interval classifiers. 
To show this, let us consider a set of labels $y_1, \dots, y_{\ell+1} \in \{ 0, 1\}$. Clearly, all points can be mapped to the label $1$ by an element of the hypothesis class. Further we can assume without loss of generality that $y_i = 0$ for at least one element $i$ and, by the periodicity of the classifiers, we can assume $i = 1$. Therefore, $\{x_1, \dots, x_{2n+1}\}$ can be shattered by periodic interval classifiers on $[a,b]$ if $\{x_2, \cdots, x_{2n+1}\}$ can be shattered by non-periodic interval classifiers on $[(x_1 + x_2)/2, b]$. We assume again, without loss of generality, that $x_1 < x_2 < \cdots$. This follows by{mohriFoundationsMachineLearning}, showing that the VC dimension of the periodic interval classifier is at least $2n+1$. 
The proof is completed by observing that for any $\{x_1, \dots, x_{2n+2}\}$ the alternating label sequence $\{0,1, \dots, 0,1\}$ cannot be produced using $n$ periodic interval classifiers. 

The same analysis can be performed with higher-dimensional input data, also leading to a finite VC dimension, that is growing with the amount of layers (see Supplementary Material note I). Hence, from a theoretical point of view, the original scheme, implemented in our work, is proven to generalize to unknown data. Let us note, however, that the finiteness of the VC dimension does not imply a non-universality of the learning model.

To gauge the effectiveness with which the algorithm can find the optimal parameters during training, let us consider the smoothness of the loss landscape. Comparing the two schemes, visually represented in Fig.~\ref{fig:vis_VCdim}, we can notice a drastic difference in the appeal of the two loss landscapes indicating possible problems for both trainability and generalizability of the compressed one. Indeed, learning models with flatter minima, like the original scheme, tend to generalize better to new data and tend to be less sensitive to noise \cite{hochreiterFlatMinima1997,keskarLargeBatchTrainingDeep2016,chaudhariEntropySGDBiasingGradient2019}. 
This is tightly related to the \textit{sharpness} of the minima, namely how quickly the loss function changes around them, which can be measured by its curvature (second derivative or Hessian). This can quantify the difficulty of landing in such a minimum, since it requires carefully balancing the learning rate: too large may cause overshooting, while too small leads to slow progress.
If we reconsiders the worst-case data set, which was used to demonstrate that the VC dimension of the compressed model is indeed infinite, we can numerically determine the sharpness of the minima in the two situations, in addition to visually interpreting the loss landscape. This is, as previously mentioned, related to the second derivative (Hessian) of the loss function. Note that since the model is not guaranteed to converge to the optimal minimum, the sharpness can only be evaluated in terms of an interval. In this context, its lower bound can be quantified by the largest eigenvalue of the Hessian matrix at the minimum itself \cite{marionDeepLinearNetworks2024}. In our case, our numerical approximation gave $\approx 0.23$ for the original scheme and $\approx 7.23 \cdot 10^{12}$ for the compressed scheme. This thereby suggests drastically sharper minima for the compressed scheme.


To summarize, our model form, which separates the encoding and processing parts, is mathematically guaranteed to be an effective learner (according to the PAC learning definition). This implies that an arbitrary lower bound on the error rate on a test set can always be reached, with a finite number of samples. This is not guaranteed to happen in the compressed scheme. Let us anyway point out that both learning models are indeed universal approximators \cite{perez2024universal, Perez-Salinas_2021} (see Supplementary Material note II), while differing drastically in their respective generalization and trainability properties.

\section{Experimental Apparatus}

The experimental implementation of the data re-uploading described in the previous sections was achieved with the experimental apparatus shown in Fig.~\ref{fig:experimental_apparatus}. This can be split into three parts: the input generation, the quantum processing and the classical post-processing.  

The photonic input was generated by a collinear Type-II spontaneous parametric down conversion source, pumped at 775~nm, emitting degenerate photons pairs at a wavelength of 1550~nm. The non-linear crystal that was used is a periodically poled titanyl phosphate, placed in a Sagnac interferometer. The crystal was pumped only in one spatial direction to generate the separable state $|01\rangle$. Then, one photon was only used to herald the generation of the second one that was implementing the qubit state. 


 For the qubit encoding, we adopt the dual rail scheme, i.e. considering two spatial modes, the presence of one photon (or of the coherent light) on the first mode would implement the state $|0\rangle$ and, on the second, the state $|1\rangle$.

 Then the quantum processing was performed through a six mode universal photonic integrated circuit\cite{clements2016optimal}, realized via femtosecond laser writing and equipped with thermo-optic phase shifters, for implementing the required unitary transformations \cite{ceccarelliLowPowerReconfigurability2020}. The input and output ports are directly interfaced with single mode fibers, which are pigtailed to the circuit. Although the employed circuit features $6$ input and outputs, we only used a $2 \times 2$ submatrix of the whole 6-mode unitary transformation, corresponding to a single MZI. Indeed, considering the conceptual scheme in Fig.~\ref{fig:conc_scheme}, we can evaluate the overall unitary operation of the circuit, in Eq.~\eqref{eq:Data_re-uploading} and implement it on a single interferometer.

 In the end, the two employed output modes of the photonic processor are connected to superconducting nanowire single photon detectors.
The frequencies of the clicks registered by the two detectors at the output of the chip, i.e. $D_2$ and $D_3$, in coincidence with the one detecting the heralding photon ($D_1$), give the probabilities $p_2$ and $p_3$ as $p_i=\frac{N_{i,a}}{\sum_{j=1}^2 N_{j,a}}$.

The output of the circuit, i.e. the variable $p_2$, for all of the different inputs, is then fed into a LDA, which finds the threshold that best separates the elements belonging to the two classes.

At the beginning of the procedure, the trainable parameters, i.e. the rotations that are in between the encoding unitaries, are set to random values. Then, during the training phase, these parameters are updated to enhance the accuracy in the reproduction of the training data. This training can be performed via numerical simulations, by employing a gradient descent algorithm \cite{kingmaAdamMethodStochastic2017} or directly on the quantum platform, where gradients are approximated through a parameter shift rule \cite{Schuld_2019} (see Supplementary Material note III).

In the end, after the retrieve of the optimal parameters (i.e. those which minimize the classification error on the training data) we can check the accuracy in the predictions of the model on new unknown data (namely, the test set). This phase is performed by encoding the test data into the circuit and, analogously to before, collecting the output statistics and feeding it into the LDA.

\section{Results}

We tested our photonic data re-uploading scheme on four different tasks, of increasing complexity. These correspond to a particular dataset, whose elements we want to classify: (i) the \textit{circles}, (ii) the \textit{moons}, (iii) the \textit{tetrominos} and (iv) the \textit{Overhead MNIST}.

The \textit{circles} and \textit{moons} dataset feature bi-dimensional data. The former is created using the implementation of the Python library \textit{scikit-learn} \cite{pedregosa2011scikit} . Hereby the scaling factor between the inner and outer circles is $0.6$ with $0.05$ Gaussian noise. The second dataset, instead, features two interleaving half circles. The points lying on the first curve are labelled as $1$ and the ones lying on the second are labelled as $2$. In order to generate this dataset, again the implementation from \textit{scikit-learn} \cite{pedregosa2011scikit} was used, with $0.1$ Gaussian noise. In both cases, the same setting, including the noise level, are used for both training and testing data. These data points, being bidimensional, can be encoded by a single MZI. Hence, in this case, the base layer block is constituted by two interferometers (one for encoding and one with trainable parameters). In Fig.~\ref{fig:results}a and Fig.~\ref{fig:results}b, we show the results for the training and test accuracy and we can see how 3 layers are sufficient to achieve a perfect classification.

The tetromino dataset, instead, features $3\times3$ pixel figures, representing ``T" and ``L" letters, corresponding to classes $1$ and $2$, as in Fig.~\ref{fig:results}c.
These letters are represented by black and white pixels. Then, to make the task more complex, we add uniform background noise bounded between $-0.1$ and $0.1$ to these 3 by 3 matrices and we add to the dataset also the negative of the images. These data points contain 9 binary values and hence require five MZIs to be encoded. Therefore, the base layer block will display 10 MZIs, i.e. five times the sequence of encoding and training. In Fig.~\ref{fig:results}c, we show the results for the training and test accuracy. In this case, we can see that the test accuracy is reduced, for a too high number of layers. Indeed, an excessive complexity of the model leads to overfitting the dataset.

For the Overhead MNIST dataset \cite{noeverOverheadMNISTBenchmark2021}, we considered the task of classifying between ships and cars. In this case, the images were pre-processed through principal component analysis  \cite{jolliffePrincipalComponentAnalysis1986,Pearson01111901} featuring 20 parameters. Hence, in this case, the base layer block was composed by 20 interferometers. Given the very high amount of optimizable parameters, we performed the experiment only up to 4 layers and the corresponding accuracy is shown in Fig.~\ref{fig:results}d.

The uncertainty was evaluated considering the effect of finite photon statistics, which follows the Poisson distribution. Hence, the error bars plotted in Fig.~\ref{fig:results} show the standard deviation obtained in the accuracy of the model, considering photon counting fluctuations. It can be seen that increasing the number of layers make our model more robust to this type of noise.

\section{Discussion}
In this study, we have demonstrated the implementation of a data re-uploading scheme on an integrated photonic platform, achieving high accuracies in several image classification tasks of increasing complexity. Our results provide both experimental and theoretical insights into the capabilities of data re-uploading, particulary in terms of the implementation of that algorithm and the influence on trainability and generalizability.

Our experimental results show that with a minimum of resources, i.e. one qubit states and a sequence of MZIs, the data re-uploading scheme can be effectively used for binary image classification tasks. The theoretical analysis complements these findings by proving that our implementation is not only a universal approximator but also an effective learner, offering a smoother optimization landscape, improved noise stability and most importantly, provable generalizability. These properties underscore the potential for practical applications of this data re-uploading model, both as a stand-alone classifier, as well as a sub-routine in more complex protocols \cite{rodriguez-grasaTrainingEmbeddingQuantum2024,freinbergerQuantumclassicalReinforcementLearning2024}.

Compared to previous implementations, our approach maintains a clear separation between encoding and processing gates. This distinction is crucial as it enhances the trainability and generalizabilty of the learning model. Indeed, as we show, compressed implementations of the data re-uploading can bring issues like overfitting, instability and a lack of provable generalization properties, due to their infinite VC dimension.

Despite the promising results, the above statements about trainabilty and generalizabilty for the compressed scheme are worst-case examples and might not be as dramatic in other tasks. However, since the data re-uploading protocol is also often used as a subroutine in other machine learning algorithms one must consider separating encoding and processing gates, as advised in this paper, to circumvent foreseeable issues, being during training or when evaluating generalization performance. Future research should focus on investigating the exact effect if data re-uploading is used as a subroutine. Additionally, from the experimental side of this work, one should focus on scaling photonic data re-uploading models to handle more complex and higher-dimensional data. This likely includes expanding to larger photonic systems and continue to investigate training on the photonic hardware \cite{facelliExactGradientsLinear2024, Schuld_2019}, as it was already done for one qubit in this paper. Further, one could study regularization schemes in order to also have provable generalization capacities for the compressed scheme.

To conclude, our study demonstrates the potential of data re-uploading schemes in quantum machine learning, particularly when implemented on photonic platforms. The findings highlight the importance of maintaining a separation between encoding and processing gates to ensure model generalizability and trainability. Furthermore, the optical implementation lays the groundwork for future energy-efficient machine learning models, whether quantum-inspired or fully quantum, for practical applications.

\section{Acknowledgements}
MFXM thanks Adrián Pérez-Salinas for a valuable discussion.
The computational results presented have been achieved in part using the Vienna Scientific Cluster (VSC). This work was supported by the Austrian Research Promotion Agency (FFG) [PIQLearn ID: 54460501]. This research was funded in part by the Austrian Science Fund (FWF)[10.55776/F71] (BeyondC) and [10.55776/I6002] (PhoMemtor). For open access purposes, the author has applied a CC BY public copyright license to any author accepted manuscript version arising from this submission. Co-funded by the European Union (HORIZON Europe Research and Innovation Programme, EPIQUE, No 101135288). Views and opinions expressed are however those of the author(s) only and do not necessarily reflect those of the European Union or the European Commission-EU. Neither the European Union nor the granting authority can be held responsible for them. MFXM wants to acknowledge the financial support by the Vienna Doctoral School in Physics (VDSP). The integrated photonic processor was partially fabricated at PoliFAB, the micro- and nanofabrication facility of Politecnico di Milano (https://www.polifab.polimi.it/). R.A., F.C. and R.O. wish to thank the PoliFAB staff for their valuable technical support.

\end{document}


\title{Supplementary Material}

\maketitle
\section{VC dimension of quantum learning models}
In order to determine the VC dimension of the investigated quantum learning models, i.e. the compressed and original data re-uploading scheme, one can try to find a simpler approximation of those systems. In that way, we will find an upper bound. If one is interested in a lower bound of the VC dimension of quantum circuits the reader is referred to \cite{Chen_2022}.
\subsection{Compressed scheme}
As mentioned in the main text, each operation on the quantum system will be implemented via a Mach-Zehnder interferometer (MZI), whose action is reported in Eq.~(3) of the main text. Hereby, we simplify our notation by setting $\phi^E_2 = 0$ and writing $\phi^E_1 = \phi$ for the external phase-shifter. Thereby the Jones matrix simplifies to:
\begin{equation}
    \label{eq:simple_MZI}
    ie^{i\frac{\theta}{2}} \begin{pmatrix}
        e^{i\phi} \sin{\left(\frac{\theta}{2}\right)} & e^{i\phi} \cos{\left(\frac{\theta}{2}\right)} \\
        \cos{\left( \frac{\theta}{2}\right)} & - \sin{\left(\frac{\theta}{2}\right)}
    \end{pmatrix}
\end{equation}

In order to demonstrate that the compressed scheme does indeed have an infinite VC dimension, we just need to consider the action of the quantum system on one fixed input state and a single layer of MZIs. If we insert the state $\ket{1}$, where $\phi = 0$ and $\theta = \omega \cdot x$ the system evolves as follows:
\begin{equation}
    \label{eq:compressed_scheme}
    ie^{i\frac{\omega x}{2}} \begin{pmatrix}
        \sin{\left(\frac{\omega x}{2}\right)} & \cos{\left(\frac{\omega x}{2}\right)} \\
        \cos{\left( \frac{\omega x}{2}\right)} & - \sin{\left(\frac{\omega x}{2}\right)}
    \end{pmatrix}
    \begin{pmatrix}
        0 \\ 1
    \end{pmatrix} =
    ie^{i\frac{\omega x}{2}}
    \begin{pmatrix}
        \cos{\left( \frac{\omega x}{2}\right)}\\
        - \sin{\left( \frac{\omega x}{2}\right)}
    \end{pmatrix}
\end{equation}
Since our measurements are insensitive to global phases, the probability of finding the system in the state $\ket{1}$ after the operation is $p_1 = |-\sin{\left( \frac{\omega x}{2}\right)}|^2$ and in the state $\ket{0}$ is $p_0 = |\cos{\left( \frac{\omega x}{2}\right)}|^2$. Typically, one assigns a point $x$ to class 0 if $p_0 > p_1$ and to class 1 otherwise. Furthermore, one can simplify $p_0 = \frac{1}{2} (\cos{(\omega x )} + 1)$ and $p_1 = \frac{1}{2}(1 - \cos{(\omega x)})$. For that reason, we can assign the class based on the sign of $p_0 - p_1 = \cos{(\omega x)}$. Therefore, our set of implementable functions is given by:
\begin{equation}
    \mathcal{H} = \left\{x \mapsto \sin{\left(\omega x + \frac{\pi}{2}\right)}| \omega \in \mathbb{R}\right\}.
\end{equation}

Since we know that the VC dimension of the set of sinusoidal functions is infinite \cite[Example 3.16]{mohriFoundationsMachineLearning}, the VC dimension of the compressed scheme is also infinite. 

\subsection{Original scheme}

In order to estimate the VC dimension of the original scheme, it is more convenient to work with the following (equivalent) Jones matrix of a MZI: 
\begin{equation}
    \frac{1}{2}\begin{pmatrix}
    \left( -1 + e^{i\theta}\right) e^{i\phi} & i\left(1 + e^{i\theta}\right)\\
    i\left( 1 + e^{i\theta}\right) e^{i\phi} & \left(1 - e^{i\theta} \right)
\end{pmatrix}.
\end{equation}

To recap, the action of a single layer in the original scheme $U_L(\vec{x}, \vec{\vartheta})$ is given by two MZIs, one depending only on the data ($\vec{x}$), the other solely on tuneable parameters ($\vec{\vartheta}$), one after the other:
\begin{align*}
    U_L(\vec{x},\vec{\vartheta}) &= \frac{1}{4}\begin{pmatrix}
    \left( -1 + e^{i\vartheta_2}\right) e^{i\vartheta_1} & i\left(1 + e^{i\vartheta_2}\right)\\
    i\left( 1 + e^{i\vartheta_2}\right) e^{i\vartheta_1} & \left(1 - e^{i\vartheta_2} \right)\end{pmatrix}
    \begin{pmatrix}
    \left( -1 + e^{ix_1}\right) e^{ix_2} & i\left(1 + e^{ix_1}\right)\\
    i\left( 1 + e^{ix_1}\right) e^{ix_2} & \left(1 - e^{ix_1} \right)\end{pmatrix}\\
    &=  \begin{pmatrix}
        W_{-+--}(\vec{\vartheta}) e^{ix_2} + W_{---+}(\vec{\vartheta}) e^{i(x_1+x_2)} & i W_{+-++}(\vec{\vartheta}) + i W_{---+}(\vec{\vartheta}) e^{ix_1}\\
        i W_{+---}(\vec{\vartheta}) e^{ix_2} + i W_{++-+} (\vec{\vartheta}) e^{i(x_1 + x_2)} & W_{+---} (\vec{\vartheta}) + W_{--+-} (\vec{\vartheta}) e^{ix_1}
   \end{pmatrix}\\
   \textnormal{\hspace{-2cm}with:}\\
   W_{\pm\pm\pm\pm} (\vec{\vartheta}) &= \frac{1}{4} \left( \pm 1 \pm e^{i\vartheta_1} \pm e^{i\vartheta_2} \pm e^{i(\vartheta_1 + \vartheta_2)}\right)
\end{align*}

In the simplest example, we evaluate the learning system with one layer and one dimensional data $x_1 = x, ~~x_2 = 0$. Hereby we will again insert the state $\ket{1}$:
\begin{equation}
    \begin{pmatrix}
        W_{-+--}(\vec{\vartheta}) + W_{---+}(\vec{\vartheta}) e^{ix} & i W_{+-++}(\vec{\vartheta}) + i W_{---+}(\vec{\vartheta}) e^{ix}\\
        i W_{+---}(\vec{\vartheta})  + i W_{++-+} (\vec{\vartheta}) e^{ix} & W_{+---} (\vec{\vartheta}) + W_{--+-} (\vec{\vartheta}) e^{ix}
   \end{pmatrix}
   \begin{pmatrix}
       0 \\ 1
   \end{pmatrix} = 
   \begin{pmatrix}
       i W_{+-++}(\vec{\vartheta}) + i W_{---+}(\vec{\vartheta}) e^{ix}\\
       W_{+---} (\vec{\vartheta}) + W_{--+-} (\vec{\vartheta}) e^{ix}
   \end{pmatrix}
\end{equation}
Therefore, the probabilities for the measured state after the operation is given by: 
\begin{align}
    p_0 &= |i W_{+-++}(\vec{\vartheta}) + i W_{---+}(\vec{\vartheta}) e^{ix}|^2\\
    p_1 &= |W_{+---} (\vec{\vartheta}) + W_{--+-} (\vec{\vartheta}) e^{ix}|^2.
\end{align}
Similar to the derivation in the previous subsection, we are using the sign of $p_0 - p_1$ to assign the class label. However, in addition, we introduce a free parameter $t$ which is given by the threshold found by the Linear Discriminant Analysis (LDA) to maximize the classification accuracy. This yields the following set of implementable functions:
\begin{equation}
    \mathcal{H} = \left\{ x \mapsto \text{sign} [a + b\cos{x} + c\sin{x}] | a, b, c \in \mathbb{R}\right\}
\end{equation}
with
\begin{align}
    a &= |W_{+-++}(\vec{\vartheta})|^2 + |W_{---+}(\vec{\vartheta})|^2 - |W_{+---}(\vec{\vartheta})|^2 - |W_{--+-}(\vec{\vartheta})|^2 + t\\
    b &= 2\Re{(W_{+-++}(\vec{\vartheta}) W_{---+}^*(\vec{\vartheta}))} - 2\Re{(W_{+---}(\vec{\vartheta})W_{--+-}^*(\vec{\vartheta}))}\\
    c &= 2\Im{(W_{+-++}^*(\vec{\vartheta})W_{---+}(\vec{\vartheta}))} - 2 \Im{(W_{+---}^*(\vec{\vartheta})W_{--+-}(\vec{\vartheta}))}.
\end{align}
Since we are mainly interested in an upper bound for the VC dimension, we can assume that $a,b, \text{ and } c$ are free tunable parameters. Furthermore these parameters can be simplified to the following:
\begin{align}
    a &= t \\
    b &= -\cos{(\vartheta_2)}\\
    c &= \frac{1}{2} \left(\sin{(\vartheta_1 + \vartheta_2)} - \sin{(\vartheta_1 - \vartheta_2)}\right)
\end{align}
Further, one can rewrite the evaluation of the VC dimension $$a + b\cos{x} + c\sin{x} =  a + R \cos{(x - \varphi)}$$ with $$R = \sqrt{b^2 + c^2} \text{~and~} \varphi=\arctan \left(\frac{c}{b}\right).$$
Assuming that $|a| \leq R$, we can identify the $2\pi$ periodic transition points of the sign of the function: $$\alpha = \varphi +\arccos \left(-\frac{a}{R}\right) \text{~ or ~} \beta = \varphi + 2\pi - \arccos \left(-\frac{a}{R}\right)$$
Therefore, the sign of the function is positive for $x \in (\alpha, \beta) \mod 2\pi$ and negative for $x \in (\beta, \alpha + 2\pi) \mod 2\pi$.
One can now write the set of implementable functions as follows:
\begin{equation}
    \mathcal{H} = \{x \mapsto \mathds{1}_{\mathcal{S}=[\alpha + 2\pi n, \beta + 2\pi n]} | \alpha,\beta \in \mathbb{R}; \forall n \in\mathbb{Z}\}.
\end{equation}
This implies that all the points in the interval \mbox{$[a + 2\pi n, b + 2\pi n]$} are classified as belonging to class 1 and those outside to class 2.
In contrast if $|a| > R$, a constant function is implemented.
It is easy to see that for such a system the VC dimension is 3, since we already know that the VC dimension of a single one-dimensional interval is 2 \cite[Example 3.11]{mohriFoundationsMachineLearning} and the periodicity of the system extents the VC dimension by one. A detailed proof of this (as well as the extension to more layers) can be found in the main text.

The same analysis can be extended to multiple layers, either still with one-dimensional inputs $x$ or generalized to higher dimensional input data. For simplicity, we will limit the data encoding to be one-dimensional for each encoding MZI, but note that behavior of the learning system is similar when using both phases of the MZI to encode data. Note that this statement is true for every layer except for the very first, where the apparatus is intrinsically insensitive to the external phase of the MZI, and thereby to one of the two encoded phases - the one denoted $x_2$ in $U_L(\vec{x},\vec{\vartheta})$.

The quantum learning system with two layers can therefore be written as:
\begin{align*}
    U_L(x_2,\vec{\vartheta_2}) U_L(x_1,\vec{\vartheta_1}) = 
    &\begin{pmatrix}
        W_{-+--}(\vec{\vartheta_2}) + W_{---+}(\vec{\vartheta_2}) e^{ix_2} & i W_{+-++}(\vec{\vartheta_2}) + i W_{---+}(\vec{\vartheta_2}) e^{ix_2}\\
        i W_{+---}(\vec{\vartheta_2})  + i W_{++-+} (\vec{\vartheta_2}) e^{ix_2} & W_{+---} (\vec{\vartheta_2}) + W_{--+-} (\vec{\vartheta_2}) e^{ix_2}
    \end{pmatrix} \cdot\\
    &\begin{pmatrix}
        W_{-+--}(\vec{\vartheta_1}) + W_{---+}(\vec{\vartheta_1}) e^{ix_1} & i W_{+-++}(\vec{\vartheta_1}) + i W_{---+}(\vec{\vartheta_1}) e^{ix_1}\\
        i W_{+---}(\vec{\vartheta_1})  + i W_{++-+} (\vec{\vartheta_1}) e^{ix_1} & W_{+---} (\vec{\vartheta_1}) + W_{--+-} (\vec{\vartheta_1}) e^{ix_1}
    \end{pmatrix}
\end{align*}
Therefore, the probabilities for the measured state after the operation is given by:
\begin{align*}
    p_0 = |&( W_{-+--}(\vec{\vartheta_2}) + W_{---+}(\vec{\vartheta_2}) e^{ix_2})(i W_{+-++}(\vec{\vartheta_1}) + i W_{---+}(\vec{\vartheta_1}) e^{ix_1}) +\\ &(i W_{+-++}(\vec{\vartheta_2}) + i W_{---+}(\vec{\vartheta_2}) e^{ix_2})(W_{+---} (\vec{\vartheta_1}) + W_{--+-} (\vec{\vartheta_1}) e^{ix_1})|^2 \\
    p_1 = |&(i W_{+---}(\vec{\vartheta_2})  + i W_{++-+} (\vec{\vartheta_2}) e^{ix_2})(i W_{+-++}(\vec{\vartheta_1}) + i W_{---+}(\vec{\vartheta_1}) e^{ix_1}) +\\
    &(W_{+---} (\vec{\vartheta_2}) + W_{--+-} (\vec{\vartheta_2}) e^{ix_2})(W_{+---} (\vec{\vartheta_1}) + W_{--+-} (\vec{\vartheta_1}) e^{ix_1})|^2
\end{align*}
If both layers share the same data input, i.e. $x_1 = x_2 = x$, then the system can approximated by the following set of implementable functions:
\begin{equation}
    \mathcal{H} = \{x \mapsto a + b \cos{(x)} + c \sin{(x)} + d \cos{(2x) + e \sin(2x)} | a,b,c,d,e \in \mathbb{R}\}
\end{equation}
with additional constraints on the parameters ($a,b,c,d$ and, $e$), which areimplied by the particular implementation of the system. The upper bound of the VC dimension of the learning system, as mentioned in the main text, therefore increases from dimension 3 to dimension 5. Thereby, one can see that the upper bond of the VC dimension, as also hypothesized by the experimental results, grows with increasing amount of layers.

If, however, the two layers do not share the same input data, in order to implement higher-dimensional data, the system can be approximately described by the following set of implementable functions with constraint parameters:
\begin{equation}
    \mathcal{H} = \{(x_1, x_2) \mapsto a + b \cos{(x_1)} + c \sin{(x_1)} + d \cos{(x_2) + e \sin(x_2)} + f \cos{(x_1 + x_2)} + g \sin{(x_1 +x_2)} |~ a,b,c,d,e,f,g \in \mathbb{R}\}.
\end{equation}
The generalization to higher-dimensional input data thereby shows that the output function depends non-linearly on each of the individual input dimensions and on their sum. 

Both of these systems can again be written in the form of a periodic interval classifier in a similar manner to the simple one-dimensional case.

\section{Universal Approximation Theorem}
The universal approximator property of the original model was recently shown in \cite{perez2024universal}, analogously to the compressed scheme \cite{Perez-Salinas_2020, Goto_2021}. First, we demonstrate this by analyzing the approximated set of implementable functions. Any function that is periodic, e.g. $2\pi$-periodic, and continuous can be represented in terms of the Fourier series, as follows:
\begin{equation}
    g(x) = \frac{a_0}{2} + \sum_{k=1}^\infty \left[a_k \cos{(kx)} + b_k \sin{(kx)}\right].
\end{equation}
Furthermore, we also know from the best approximation properties of the Fourier series, see for example \cite[Lemma 1.2]{steinFourierAnalysisIntroduction2003}, that the truncated Fourier series yields the best possible approximation of any periodic square integrable function. Trigonometric polynomials are, therefore, dense in the space of continuous periodic functions with respect to the uniform norm.
Similarly, the general expression of the $N$-layered learning model in the derived approximation picture is:
\begin{equation}
    \mathcal{H}=\left\{x \mapsto a +\sum_{l=1}^N \left[        b_l \cos{(lx)} + c_l \sin{(lx)}\right] ~\middle|~ a,b_l, c_l \in \mathbb{R}\right\}
\end{equation}
By allowing $N \rightarrow \infty$, the learning model is equivalent to the Fourier series representation of any continuous periodic function. Thereby, also the original implementation of the learning model is a universal approximation for continuous functions on a compact interval. 

\section{On-chip and numerical training Procedure}
The models shown in this publications were all trained in a similar fashion. Hereby, we mainly utilised the python libraries TensorFlow \cite{developers2022tensorflow} and Strawberry Fields \cite{Killoran_2019}. We implemented a train of universal Mach-Zehnder Interferometers (MZIs) to encode the data and the tunable parameters and simulated the propagation with single photons, or equivalently coherent light, through the system. In the training process, we used the Adam optimizer as implemented by TensorFlow with default parameters. As a loss function a linear discriminant analysis (LDA) loss \cite{dorferDeepLinearDiscriminant2016} was used and a LDA was used for assigning the class labels. The entire system was then trained to convergence, for a maximum of $10.000$ iterations. The necessary gradients were calculated via automatic differentiation. 

Furthermore, we also investigated the training on the experimental hardware. This was implemented via the parameter shift rule \cite{Schuld_2019}. 
In this context, we also investigated the performance of a "forward", "backward", and "central" implementation. We found out that, for the investigated datasets, a combination of "backward" finite differences and "central" finite difference, yielded the best ratio of experimental runtime (number of runs per gradient estimation) and training performance.
While the architecture was kept unchanged, except for the derivation of the gradients, the learning rate was adapted the yield faster convergence in order to ensure experimental feasibility.